**Formation of Double Neutron Stars, Millisecond pulsars and Double Black Holes**

Edward P.J. van den Heuvel

Anton Pannekoek Institute of Astronomy, University of Amsterdam, Science Park 904, 1098XH Amsterdam, The Netherlands, and

Kavli Institute for Theoretical Physics, University of California Santa Barbara, CA 93106-4030, USA

**Abstract**

The 1982 model for the formation of Hulse-Taylor binary radio pulsar PSR B1913+16 is described, which since has become the "standard model" for the formation of the double neutron stars, confirmed by the 2003 discovery of the double pulsar system PSR J0737-3039AB. A brief overview is given of the present status of our knowledge of the double neutron stars, of which 15 systems are presently known. The binary-recycling model for the formation of millisecond pulsars is described, as put forward independently by Alpar et al. (1982), Radhakrishnan and Srinivasan (1982), and Fabian et al. (1983). This now is the "standard model" for the formation of these objects, confirmed by the discovery in 1998 of the accreting millisecond X-ray pulsars. It is noticed that the formation process of close double black holes has analogies to that of close double neutron stars, extended to binaries with larger initial component masses, although there are also considerable differences in the physics of the binary evolution at these larger masses.

1. Introduction

The X-ray binaries were discovered in 1972 (Schreier et al. 1972) and the first binary radio pulsar was discovered in 1974: the Hulse-Taylor pulsar PSR B1913+16, in a very eccentric binary system (e=0.617) with a very short orbital period (7h45 minutes; Hulse and Taylor 1975). Although soon after the discovery of this system it was realized that it most probably originated from a High-Mass X-ray Binary (Flannery and van den Heuvel 1975, De Loore et al. 1975), there were still many puzzling questions concerning its precise evolutionary history. Particularly its very short pulse period (59 ms) in combination with its weak dipole magnetic field (~ $10^{10}$ G) made this pulsar very anomalous. It lasted until 1980/1982 before a consistent evolutionary picture emerged, which led to the prediction that the companion of this pulsar must be a normal "garden variety" pulsar with a strong magnetic field, which was formed after the observed pulsar in the system (Srinivasan and van den Heuvel 1982). The observed pulsar PSR B 1913+16 obtained, according to this model, its rapid spin due to a history of accretion of mass with angular momentum in a binary system, as had been suggested by Smarr and Blandford (1976). (Around 1980, for such pulsars, which were spun up by accretion, Radhakrishnan coined the name "recycled pulsar"). Shortly after this 1982 paper came out, Backer et al. (1982) announced the discovery of the first millisecond radio pulsar PSR 1937+21, an object nobody had expected. Its pulse period is 20 times shorter than that of the Crab pulsar – which until then had the shortest pulse period known- and its magnetic field is some $10^4$ times weaker than that of the normal single pulsars that were then known. Although this object is single, Alpar et al. (1982), Radhakrishnan and Srinivasan (1982) and Fabian et al. (1983), independently, extended the binary "recycling" idea to this pulsar. They proposed that it descended from a Low-Mass X-ray Binary. Although at the time this extension by a very large factor of the "recycling" model seemed a very bold step, later discoveries have proven this idea to be fully right. I describe these two developments in some more detail in the next section. In section 3, I will briefly discuss the present state of our knowledge of the evolutionary history of the double neutron stars and of double black holes.

## 2. Double neutron stars and millisecond pulsars

### 2.1 Double neutron stars

In the period 1978 to 1980 Srinivasan and I profoundly discussed the possible ways in which the Hulse-Taylor binary pulsar could have been formed. We concluded that the suggestion of Smarr and Blandford (1976), that the observed pulsar in this system (PSR B 1913+16) is a relatively old neutron star, that has been spun-up by accretion in its binary system, is indeed correct, as this is the only way to explain its very anomalous position in the P-$P_{dot}$ diagram of radio pulsars (figure 1). Srinivasan argued, on theoretical grounds, that this position is excluded for a normal newborn pulsar. We further reasoned that the very narrow orbit of the system implies that the progenitor High-Mass X-ray Binary (HMXB) system must have gone through a deep spiral-in phase, which resulted in a very close binary consisting of a helium star (the helium core of the massive donor star of the HMXB) and a neutron star, as had been computed by van den Heuvel and De Loore (1973). During this spiral-in phase tidal forces will have completely circularized the orbit. The fact that the present system has a large orbital eccentricity then can be explained only if the helium star terminated its life in a supernova explosion, which proofs that the companion of PSR B1913+16 must also be a neutron star. Since this second-born neutron star cannot have undergone any accretion in this double neutron star system, we reasoned that it must be a normal strong-magnetic field "garden variety" radio pulsar, with a magnetic field strength of order $10^{12}$ G. We argued that this later-born neutron star is not observed because new-born strong-magnetic field neutron stars such as the Crab pulsar spin down rapidly and disappear into the "pulsar graveyard" within about $10^7$ years after their formation, whereas the recycled pulsar PSR B1913+16 with its weak magnetic field of ~ $10^{10}$ G spins down very slowly and takes > $10^8$ years to disappear into the "graveyard". For this reason, the first-born recycled neutron star in a double neutron star systems is expected to remain observable long after its second-born companion has stopped pulsing. We wrote down this model for the formation of the Hulse-Taylor binary pulsar in a paper which was submitted to Astronomy and Astrophysics and accepted in 1980, but published only in 1982, due an administrative error at the journal (Srinivasan and van den Heuvel 1982). Srinivasan and I were very happy with the discovery by Lyne et al. (2004) of the double pulsar system PSR J0737-3039AB, which fully confirmed this prediction from 1980. This system consists of 22.7 msec recycled pulsar with a magnetic field of $2 \times 10^9$ G, plus a "garden variety" pulsar with a 2.7735 second period and a magnetic field strength of $0.49 \times 10^{12}$ G. The orbital period of the system is 2.4 hours, and the eccentricity is e = 0.088. Clearly, the "garden variety" pulsar PSR J0737-3039B is the second-born neutron star in the system, and has already spun down to a relatively long pulse period.

Interesting in this system is that the second-born neutron star has a relatively low mass for a neutron star (1.249 solar masses) while its first-born recycled companion has a mass of 1.338 solar masses. These masses were determined from the very accurately measured General and Special Relativistic effects measured for this system (e.g. Kramer and Stairs 2008). The low eccentricity of the system indicates that the second-born neutron star received hardly any velocity kick at its birth. This, in combination with the low mass of the second-born neutron star suggests that it originated from an electron-capture collapse of a degenerate O-Ne-Mg core, rather than from the collapse of an iron core (Podsiadlowski et al. 2004, van den Heuvel 2004), since an e-capture collapse is not expected to induce a large kick velocity to the neutron star (e.g. Kitaura et al. 2006 and references therein). One sees here that double neutron stars not only are beautiful objects for testing the predictions of the General Theory of Relativity (which earned Hulse and Taylor the 1993 Nobel Prize of physics), but they also provide key information about stellar evolution and the formation mechanisms of neutron stars. In section 3, I briefly describe the 15 double neutron stars that are presently known, in relation to the 1980/1982 evolutionary picture.

### 2.2. The first millisecond pulsar

In 1982 the first millisecond radio pulsar PSR 1937+21 was discovered by Backer et al. (1982). Its discovery was thanks to the fact that Shri Kulkarni, then a graduate student at Berkeley, had built instrumentation that could detect pulse periods of order one millisecond. He and Backer used this new instrumentation on the already long-known scintillating and highly polarized radio source 4C21.53 close to the galactic plane. The source turned out to be a pulsar (PSR 1937+21) with the amazingly short period of 1.5 msec (spin frequency 642 Hz), and an amazingly small period derivative, indicating a magnetic field strength of only about $10^8$ Gauss. Although the pulse period would suggest a very young neutron star, no supernova remnant was found around it. Immediately after this discovery, Radhakrishnan and Srinivasan (1982) and Alpar et al. (1982), and somewhat later Fabian et al. (1983), independently put forward the idea that, like the Hulse-Taylor binary pulsar PSR B 1913+16, this pulsar has been recycled in a binary system. This time: not in a High-Mass X-ray Binary, where the accretion and spin-up phase lasts relatively short (at most ~ $10^6$ years), but in a Low-Mass X-ray Binary (LMXB), where it may last $10^8$ to $10^9$ years, such that a very large amount of mass and angular momentum can be fed to the neutron star. They extended here the recycling idea by a factor 30, relative to the 59 msec pulse period of PSR B1913+16. And on top of that they had to assume that somehow the companion star in the progenitor binary, that had fed the mass and angular to this neutron star, had disappeared. When I read these first two papers, my first reaction was: "this is ridiculous, as it takes the recycling idea far out of its range of applicability". However, within a year, Radhakrishnan and Srinivasan, Alpar and colleagues and Fabian and colleagues, were proven completely right, thanks to the discovery by Boriakoff et al. (1983) of the second millisecond pulsar, which indeed is in a binary system, with a low-mass helium white dwarf as a companion star. Clearly, this system is the descendant of a relatively wide Low-Mass X-ray Binary (Joss and Rappaport, 1983; Savonije, 1983; Paczynski, 1983 and Helfand et al. 1983), as this helium white dwarf is the remnant of a low-mass donor star. We now know over one hundred millisecond radio pulsars, most of them in binary systems, and the 1982 LMXB-recycling model has been proven fully correct. The most definitive proof of its correctness came with the discovery in 1998 of the first accreting millisecond X-ray pulsar in the Low-Mass X-ray Binary SAX J1808-3658 by Wijnands and van der Klis (1998), with a spin frequency of 401.0 Hz. Thanks to NASA's Rossi X-ray Timing Explorer, several tens of these accreting millisecond X-ray pulsars in LMXBs have since been discovered (e.g. see van der Klis 2000).

## 3. Recent progress in our knowledge of the double neutron stars and their relation to double black holes

3.1 Double neutron stars known to date

In recent years, many new double neutron stars have been discovered. The total known number of such systems is now 15, as listed in table 1, according to Tauris et al. (2017). Two of them are in globular star clusters and have presumably been formed by dynamical capture processes in the dense central cores of the clusters. The other 13, in the disk of the Galaxy, are expected to have been formed by binary evolution, according to the above outlined 1980/1982 model, as a later evolutionary phase of a High-Mass X-ray binary. In this model, the direct progenitor of the double neutron star was a very close system, consisting of a helium star and a neutron star, which van den Heuvel and De Loore (1973) had assumed to result from the evolution with Roche-lobe overflow, of a High-Mass X-ray binary with a relatively short orbital period of order one week. In 1975, however, Webbink (1975) and Paczynski (1976) realized, that due to the extreme mass ratio of the donor star and the neutron star in a HMXB, Roche-lobe overflow in these systems will be unstable and will lead to the formation of a Common Envelope (CE), in which the neutron star and the helium core of the post-main-sequence donor star orbit around their common center of gravity (Paczynski refers this idea to an earlier conversation with Ostriker (1973)). Due to the large friction which these objects experience in their motion inside the CE, they will rapidly spiral towards each other. It was found that the resulting loss of orbital gravitational binding energy only is sufficient to eject the Common Envelope if the initial binary system is very wide (e.g. see Taam and Sandquist 2000). Only HMXBs with orbital periods longer than about one year will therefore survive Common Envelope Evolution as close binaries consisting of helium star and a compact star. The HMXBs most suitable for producing double neutron stars are the B-emission X-ray Binaries, of which quite a number have orbital periods longer than one year. Figure 2 depicts the thus resulting possibilities for the final evolution of HMXBs with a neutron star accretor, according to Bhattacharya and van den Heuvel (1991).

In table 1 the magnetic field strengths of all known pulsars in double neutron stars are given, as derived from their spin periods and period derivatives (see Tauris et al. 2017). Two of these pulsars, PSR J0737-3039B and PSR J1906+0746 have strong magnetic fields, about $0.5 \times 10^{12}$ G. Clearly, these are non-recycled second-born neutron stars. All other ones are recycled pulsars with magnetic field strengths between $2.10^8$ and $2.10^{10}$ G. The system of PSR J1906+0746 is particularly interesting, as it is the only one in which only the second-born neutron star is seen, and not the recycled first-born one. As pointed out by Yang et al. (2017), the orbital eccentricity, orbital period and low pulsar mass in this system all closely resemble those of PSR J0737-3039B, making these pulsar binaries look almost like identical twins. As the recycled pulsar in this system will live much longer than its non-recycled companion, this unseen recycled pulsar most likely is still an active radio pulsar. The reason why we do not observe it is, as argued by Yang et al. (2017), that we are outside its pulsar beam. Due to the second supernova mass ejection, the rotation axes of the neutron stars in double neutron star systems are most probably not directed perpendicular to the orbital plane. Therefore, these rotation axes will precess around the normal to the orbital plane due to the General Relativistic de Sitter precession. This effect has already been observed in the non-recycled component the double pulsar, PSR J0737-3039B, which disappeared in 2008 due to this effect and most probably will not become observable again before 2035 (Perera et al. 2010). PSR J1906+0746 itself shows a gradual change in the shape of its pulse profile, presumably also due to this precession effect (van Leeuwen et al. 2015). The recycled companion of PSR J1906+0746 may, due to this precession effect, become observable as a pulsar in the future, when its pulsar beam precesses into the line of sight to Earth (Yang et al. 2017).

3.3 From double neutron stars to double black holes

If in the evolutionary model for producing double neutron stars, depicted in the right-hand panel of figure 2, one would have started out with a wide black-hole HMXB with a donor star more massive than about 30 solar masses, the final result would have been a close double black hole, instead of a close double neutron star. This is, in essence, the model for the formation of close double black holes put forward, for example, by Tutukov and Yungelson (1993), Lipunov et al. (1997) and Belczynski et al. (2016). One therefore expects also close double black hole binaries to be present in galaxies, and since the gravitational wave bursts produced by their mergers will be much stronger than those of merging double neutron stars, it is not so surprising that the first GW events observed by LIGO were mergers of double black holes (Lipunov et al. 1997).

Without the large detailed body of observational evidence about the formation of close double compact objects provided by the double neutrons stars, nobody would have felt confident to make any credible prediction about the formation and existence of close double black holes. The detailed evidence that the close double neutron stars have provided us about the evolution of massive binary systems, including particularly the key phase of deep spiral-in of the first-born compact star in the envelope of its companion, was therefore crucial for being able to predict that close double black holes exist in nature.

However, it is important to notice that, especially for initial binary component masses > 30 solar masses, the physics of binary evolution is not fully analogous to that of the binaries with initial component masses < 20 solar masses, for the following reasons:

- The large stellar wind mass loss that occurs for the more massive stars, but plays hardly any role < 20 solar masses; the wind-mass loss rate increases with metallicity; for solar metallicity, the large wind mass loss prevents red super giants to be formed at masses > 30 -40 solar masses. As a result, the wide binaries with red super giants that, according to figure 2, are required to survive CE-evolution as a close binary, will not exist. So, this type of CE evolution then does not occur, at least not at solar metallicity (whether very massive red super giants will occur at low metallicities is, so far, uncertain).

- The fact that in HMXBs with a black-hole accretor, instead of CE-evolution, stable mass transfer by Roche-lobe overflow will occur, if the mass ratio of donor and accretor is < 3.5 (van den Heuvel et al. 2017), and possibly even if the mass ratio is < 6 (Pavlovski et al. 2016). This means that many of the black-hole HMXBs may avoid CE-evolution completely. Their evolution with stable Roche-lobe overflow may still lead to considerable orbital shrinking, but not as strong as expected from CE-evolution, but still a sizeable fraction of the resulting double black holes will be close enough to merge within a Hubble time (van den Heuvel et al. 2017).

4. Conclusion

The research of Ganesan Srinivasan on double neutron stars and millisecond pulsars has contributed much to our understanding of the evolution of neutron stars in binary systems. It also has provided a basis, by analogy, for understanding the formation of close double black holes.

**Acknowledgements**: This research was supported in part by the National Science Foundation under Grant No. NSF PHY11-25915.

**Captions of figures and table**:

Figure 1: First figure from the paper of Srinivasan and van den Heuvel (1982): the position of PSR B1913+16 in the $P_{dot}$ vs. P diagram for the 87 radio pulsars known in 1977. Lines of constant spin-down age $P/(2P_{dot})$ are indicated.

Figure 2: The various possibilities for the final evolution of a High-Mass X-ray Binary with a neutron-star accretor. In all cases the onset of Roche-lobe overflow leads to the formation of a common envelope and the occurrence of spiral-in. (a) In systems with orbital periods less than about one year the decrease in orbital gravitational potential energy during spiral-in is most probably not sufficient to unbind and eject the common envelope, and the neutron star spirals down into the core of the companion, forming a so-called Thorne-Zytkow star, which finally ejects its envelope due to the release of nuclear energy in the layers surrounding the neutron star. (b) In systems with orbital periods longer than about one year the common envelope is ejected during spiral-in, and a close binary is left, consisting of the neutron star and the core, consisting of helium and heavier elements, of the companion star. Companions initially more massive than about 8 solar masses leave cores that will explode as a supernova, leaving an eccentric binary pulsar, or two runaway pulsars. Systems with companions less massive than about 8 solar masses leave close binaries with circular orbits and a massive white-dwarf companion, like PSR B0655+64. (Figure from Bhattacharya and van den Heuvel 1991).

Table 1: Properties of the 15 double neutron stars with published data, including a few unconfirmed candidates (after Tauris et al. (2017), which also gives the source references of the data).

Table 1:

___________________________________________________________________________________

| Radio Pulsar | Type | $P$ (ms) | $P_{dot}$ ($10^{-18}$) | $B$ ($10^9$ G) | $P_{orb}$ (days) | $e$ | $M_{Psr}$ ($M_{sun}$) | $M_{comp}$ ($M_{sun}$) | $\tau_{Grw}$ (Myr) |
|---|---|---|---|---|---|---|---|---|---|
| J0453+1559 | recycled | 45.8 | 0.186 | 0.92 | 4.072 | 0.113 | 1.559 | 1.174 | ∞ |
| J0737-3039A | recycled | 22.7 | 1.76 | 2.0 | 0.102 | 0.088 | 1.338 | 1.249 | 86 |
| J0737-3039B | young | 2773.5 | 892 | 490 | 0.102 | 0.088 | 1.249 | 1.338 | 86 |
| J1518+4904 | recycled | 40.9 | 0.0272 | 0.29 | 8.634 | 0.249 | - - | - - | ∞ |
| B1534+12 | recycled | 37.9 | 2.42 | 3.0 | 0.421 | 0.274 | 1.333 | 1.346 | 2730 |
| J1753-2240 | recycled | 95.1 | 0.970 | 2.7 | 13.638 | 0.304 | - - | - - | ∞ |
| J1755-2550 | young(?) | 315.2 | - - | - - | 9.696 | 0.089 | - - | > 0.40 | ∞ |
| J1756-2251 | recycled | 28.5 | 1.02 | 1.7 | 0.320 | 0.181 | 1.341 | 1.230 | 1660 |
| J1811-1736 | recycled | 104.2 | 0.901 | 3.0 | 18.779 | 0.828 | <1.64 | >0.93 | ∞ |
| J1829-2456 | recycled | 41.0 | 0.0525 | 0.46 | 1.176 | 0.139 | <1.38 | >1.22 | ∞ |
| J1906+0746 | young | 144.1 | 20300 | 530 | 0.166 | 0.085 | 1.291 | 1.322 | 309 |
| J1913+1102 | recycled | 27.3 | 0.161 | 0.63 | 0.206 | 0.090 | <1.84 | >1.04 | ~480 |
| B1913+16 | recycled | 59.0 | 8.63 | 7.0 | 0.323 | 0.617 | 1.440 | 1.389 | 301 |
| J1930-1852 | recycled | 185.5 | 18.0 | 18 | 45.060 | 0.399 | <1.32 | >1.30 | ∞ |
| J1807-2500B | Glob. Cl. | 4.2 | 0.0823 | 0.18 | 9.957 | 0.747 | 1.366 | 1.206 | ∞ |
| B2127+11C | Glob. Cl. | 30.5 | 4.99 | 3.8 | 0.335 | 0.681 | 1.358 | 1.354 | 217 |

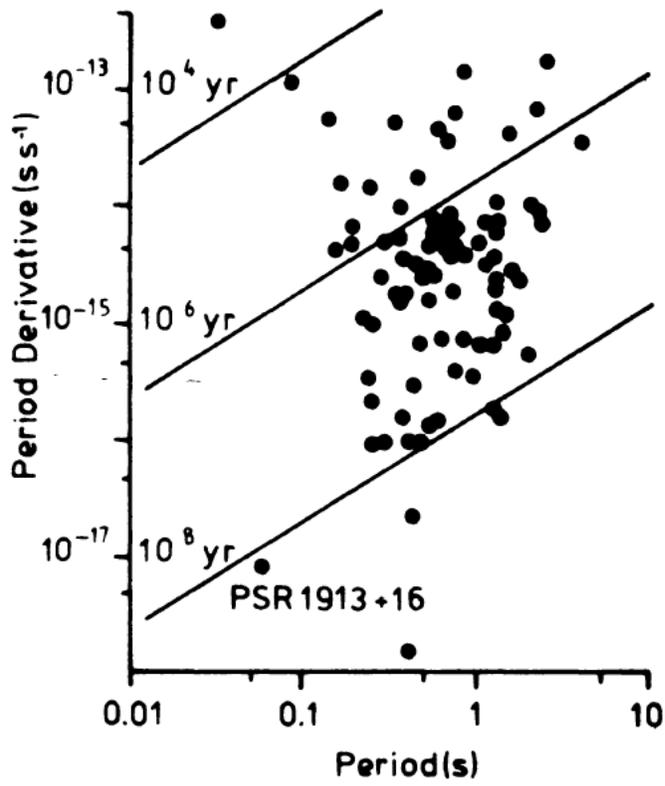

Figure 1

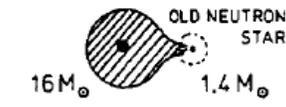
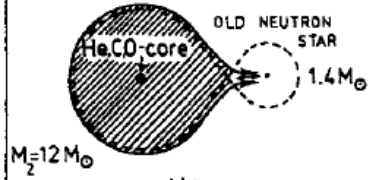
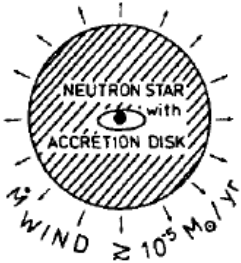
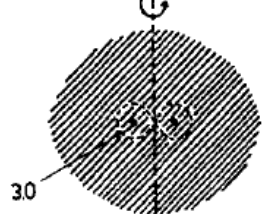
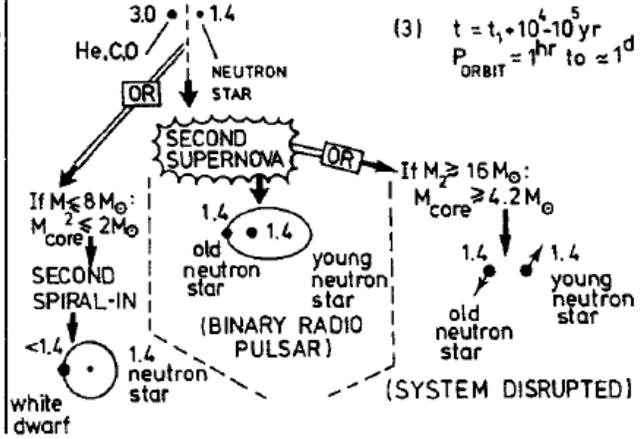

Figure 2.